\documentstyle[prl,aps,graphicx,twocolumn,floats,amstex,xspace]{revtex}

\begin{document}

\newcommand{\micron}{$\mu\mathrm{m}$\xspace}

\wideabs{
\title{Brownian Dynamics of a Sphere Between Parallel Walls}

\author{Eric R. Dufresne, David Altman, and David G. Grier}

\address{Dept. of Physics, James Franck Institute, and
Institute for Biophysical Dynamics\\
The University of Chicago, Chicago, IL 60637}

\date{\today}

\maketitle

\begin{abstract}
  We describe direct imaging measurements of a colloidal 
  sphere's diffusion between two parallel surfaces.
  The dynamics of this deceptively simple hydrodynamically
  coupled system have proved difficult to analyze.
  Comparison with approximate formulations of 
  a confined sphere's hydrodynamic
  mobility reveals good agreement with both a leading-order
  superposition approximation as well as a more general all-images
  stokeslet analysis.
\end{abstract}



} 

Stationary surfaces modify the flow field set up by
a moving particle, thereby increasing
the particle's hydrodynamic drag.
Calculating this influence
remains a vexing problem in all but the simplest geometries
because complicated
boundary conditions usually render the problem intractible.
For example,
Fax\'en derived
a single sphere's hydrodynamic coupling to
a rigid planar surface \cite{happel91} as early as 1924.
Adding a second parallel wall, however, is so much
more challenging that a definitive formulation is not yet
available.
This is particularly unsatisfactory in that many biologically
and industrially relevant processes are governed by particles' dynamics
in confined geometries.

This Letter describes measurements of a single colloidal
sphere's diffusion
through water in a slit pore formed by two parallel
glass surfaces.
We use optical tweezers \cite{grier97}
to position the sphere reproducibly within the sample volume,
a slow steady Poisseuille flow to establish its position between
the walls,
and high-resolution digital video microscopy \cite{crocker96}
to track its motions in the plane.
By positioning a test sphere at
selected heights within a slit pore, releasing it and tracking its
motions, we are able to measure its
height-dependent hydrodynamic coupling to
parallel bounding surfaces.
These measurements agree quantitatively with predictions
based on the stokeslet approximation, a tool which
is particularly useful for describing many-body
hydrodynamic interactions in colloidal suspensions
\cite{brenner99,dufresne00}.

Previous imaging \cite{faucheux94} and light-scattering \cite{lobry96}
studies probed confined spheres'
dynamics averaged
over the slit pore's thickness and
only indirectly addressed how a sphere's 
mobility changes
as it moves relative to confining walls.
A very recent study combined
digital video microscopy with optical tweezer manipulation
to measure a sphere's mobility near the midplane of a slit
pore as a function of the slit pore's width \cite{lin00}.
Its results cast
serious doubt on a recently proposed theory for confined
Brownian motion \cite{lobry96}, but left open questions
regarding a sphere's dynamics at other, less symmetric configurations.

The present measurements were performed on a single
polystyrene sulfate microsphere from a suspension of spheres
$a = 1.006 \pm 0.010$~\micron in radius (Catalog No.~4204A,
Duke Scientific) 
dispersed in an
aqueous solution of 2~mM NaCl at $T = 300.66 \pm 0.07^\circ$K.
The sample was cleaned before resuspension by extensive
dialysis against deionized water and then infiltrated into
a slit pore
of area $2~\mathrm{cm} \times 1~\mathrm{cm}$ and thickness
$H = 8$~\micron
created by sealing the edges of a \#1 coverslip to a glass microscope
slide with uv cured adhesive (Norland Type 88).
Access to the sample volume was provided by two glass tubes bonded
to holes drilled through the slide at either end of the longest dimension.
All glass surfaces were cleaned thoroughly before assembly to ensure 
uniform surface properties \cite{hair70}.
Although both the sphere and the glass walls develop large surface charges
when immersed in water, the suspension's high ionic strength
reduced the Debye-H\"uckel screening length to 7~nm and thus
minimized electrostatic interactions.
Surface separations were always large enough
that van der Waals attractions were negligible \cite{pailthorpe82}.

The sample was mounted on the stage of an Olympus IMT-2
inverted optical microscope and imaged with a $100 \times$ NA 1.4
oil immersion objective lens.
Images captured with an attached CCD camera were recorded
on a JVC BRU-S822DXU computer-controlled
SVHS video deck before being digitized with a MuTech MV-1350 frame
grabber.
Computerized analysis \cite{crocker96} of the resulting
sequence of digitized images yielded
measurements of the colloidal sphere's position in
the microscope's focal plane
with 20~nm spatial resolution
at 1/60~sec intervals \cite{crocker96}.

A colloidal sphere
diffuses through a Newtonian fluid according to
\begin{equation}
  \langle \Delta r_i^2 (\tau) \rangle = 2 D_i(\vec r) \, \tau,
  \label{eq:stokeseinstein}
\end{equation}
where $\Delta r_i(\tau)$ is its displacement in the $i$-th dimension
over time $\tau$ and $D_i(\vec r)$ is the associated diffusion coefficient.
Angle brackets indicate an ensemble average.
The fluctuation-dissipation theorem relates
$D_i(\vec r)$ to the drag force,
$\gamma_i(\vec r) \, v_i$, the sphere experiences
as it moves past point $\vec r$ with speed $v_i$:
$D_i(\vec r) = k_B T / \gamma_i(\vec r)$.
The drag coefficient, $\gamma_i$, emerges
from the solution to the
Stokes equation describing flow at low Reynolds
numbers, subject to the surfaces' no-slip boundary conditions.
\begin{figure}[t!]
  \begin{center}
    \includegraphics[width=3in]{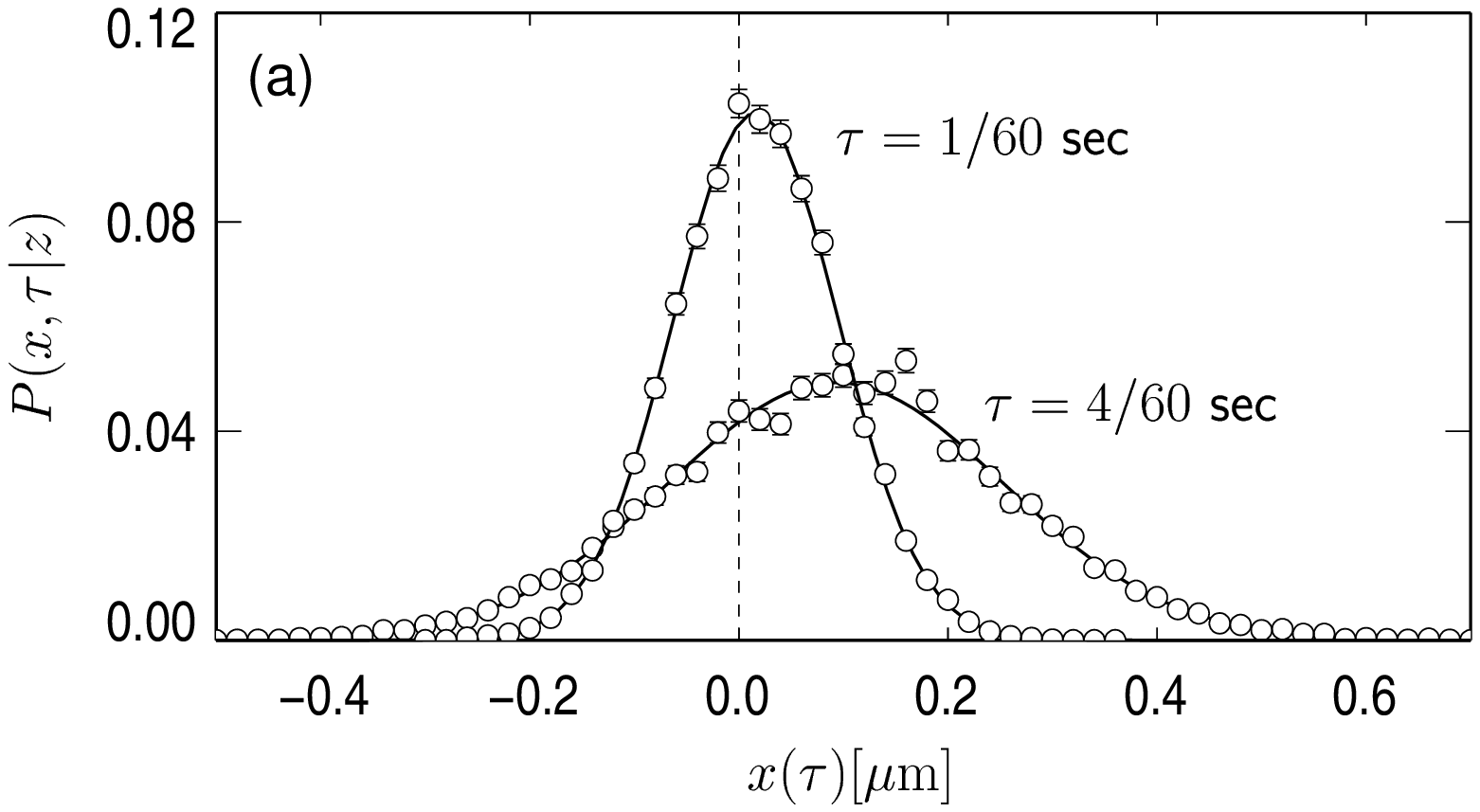}
    \includegraphics[width=1.47in]{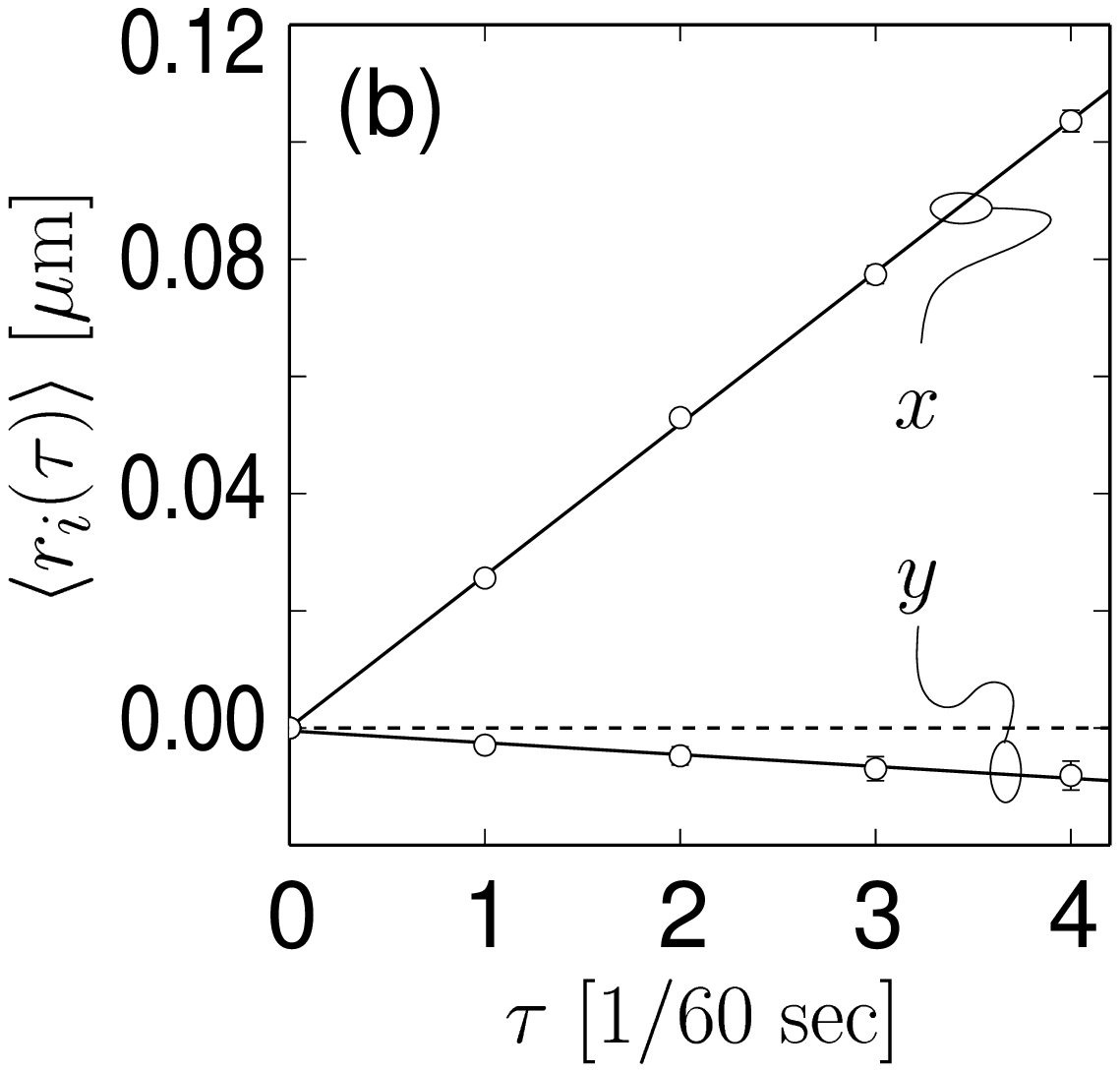}
    \includegraphics[width=1.53in]{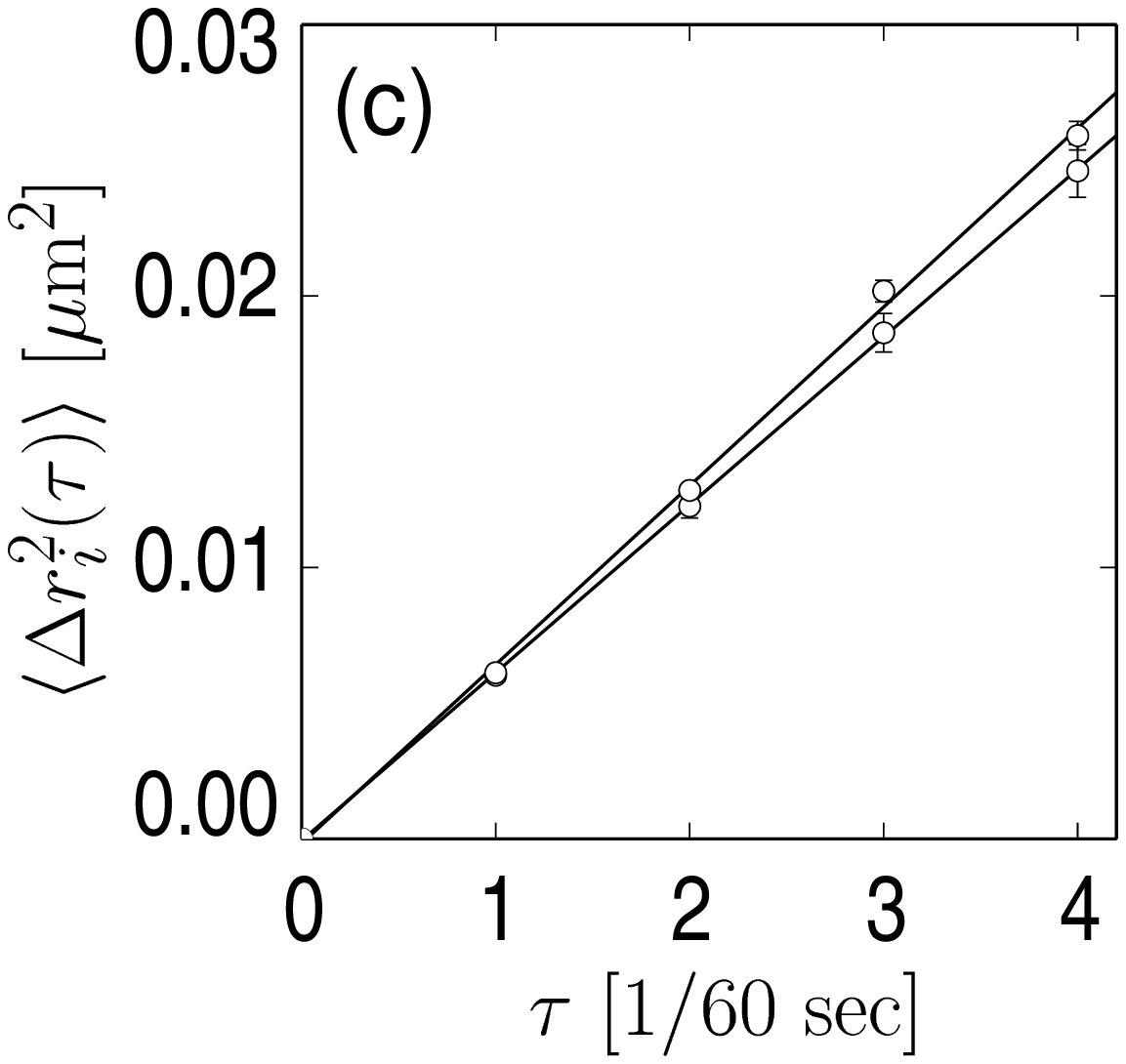}
    \caption{(a) Measured probability distribution for displacements along
      the direction of flow at $\tau = 1/60$~sec and 4/60~sec
      for $z = 4$~\micron.  Solid curves are fits to 
      Eq.~(\protect\ref{eq:Pxt})
      for the width and displacement of the distribution.
      (b) Drift of the distributions' centers along and perpendicular
      to the direction of the imposed flow.
      (c) Evolution of the mean-square widths of $P(r_i,\tau | z)$,
      fit to Eq.~(\protect\ref{eq:stokeseinstein})
      for $D_i(h)$.
      }
    \label{fig:Pxt}
  \end{center}
\end{figure}

For example, a sphere of radius $a$ suspended in
an unbounded fluid of viscosity $\eta$
has an isotropic and translation-invariant
drag coefficient, $\gamma_0 = 6 \pi \eta a$.
The corresponding free self-diffusion coefficient for the
spheres in the present study is
$D_0 = 0.259 \pm 0.04~\mathrm{\mu m^2/sec}$.
A sphere passing between two rigid parallel
walls, on the other hand,
experiences a drag which depends both on its position $z$ in
the slit and also on its direction of motion.
Measuring this dependence requires the ability either to track
a sphere's motion in three dimensions, or else to position
the sphere
reproducibly within the slit.

We adopted the second approach, using an optical tweezer to
place a single sphere in the microscope's focal plane, releasing
it for $t=5/60$~sec to measure its diffusivity, and then retrapping it.
The optical tweezer was formed by directing a collimated beam
of laser light (100~mW at 532~nm) through the objective lens'
back aperture.
Optical gradient forces exerted by the tightly converging
beam localized the sphere near the focal point despite
radiation pressure and random thermal forces.
Suddenly deflecting the beam onto a beam block with a 
galvanometer-driven mirror extinguished the optical trap and
freed the particle to diffuse \cite{dufresne00}.
Residual heating of roughly $1^\circ$C due to steady-state optical absorption
relaxed in a few microseconds through
thermal diffusion, and did not affect the
sphere's dynamics on the time scale of our data collection 
\cite{crocker96,crocker94}.
The galvanometer drive was synchronized to the video camera's sync signal
to ensure
that data acquisition began at a reproducible interval after
the optical trap was extinguished.
 
We adjusted the focal plane's height
in 1.0~\micron 
increments to sample the diffusivity's dependence on
the sphere's height in the slit pore.
The resulting uncertainty in $z$ due to the sphere's out-of-plane
diffusion was smaller than $\sqrt{2D_0 t} = 0.2$~\micron,
and so was small compared with both the sphere's diameter
and the separation between walls.

\begin{figure}[t!]
  \begin{center}
    \includegraphics[width=3in]{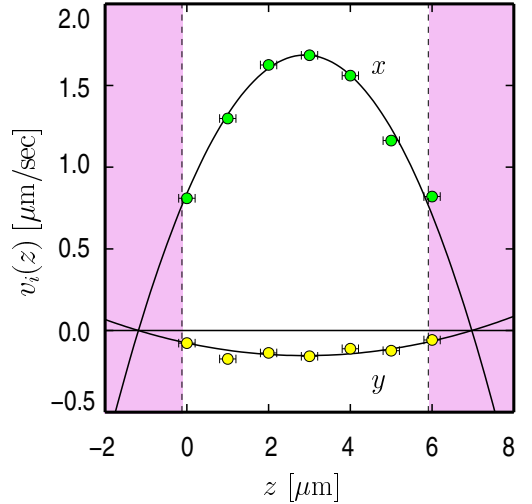}
    \caption{The drift rate's dependence on initial height in the
      slit pore.  Solid curves result from least squares fits
      to Eq.~(\protect\ref{eq:vz}).  
      Shaded regions are forbidden by sphere-wall contact,
      given the fit estimates for the wall locations.}
    \label{fig:vz}
  \end{center}
\end{figure}

We determined the absolute positions of the glass walls
by observing the influence of
a steady Poisseuille flow
on the particle's motion.
The advected sphere attained a height-dependent
drift velocity
\begin{equation}
  v_i(z) = \frac{4 v_{0,i}}{H^2} \, (z - z_0) (z_0 + H - z).
  \label{eq:vz}
\end{equation}
where $z_0$ is the position of the lower wall.
This drift caused the probability distribution for finding
the sphere at position $\vec r$ to shift as it broadened with time:
\begin{equation}
  P(r_i, \tau | z ) = \sqrt{
    \frac{1}{2 \pi \langle \Delta r_i^2(\tau) \rangle}} \,
  \exp \left( - \frac{[r_i - \langle r_i (\tau) \rangle]^2}{%
      2 \, \langle \Delta r_i^2(\tau) \rangle} \right),
  \label{eq:Pxt}
\end{equation}
where $\langle r_i(\tau) \rangle = v_i(z) \tau $ is the mean displacement in
the $i$-th direction after time $\tau$ given the 
steady flow in that direction 
at height $z$.

At each tweezer height $z$, we collected 3,000 sequences of 5 video
fields, yielding  12,000 measurements of $P(x,\tau | z)$ and $P(y,\tau | z)$
over four 1/60 second intervals.
Typical examples at the midplane of the slit pore appear
in Fig.~\ref{fig:Pxt}(a).
The peaks' drifts, plotted in Fig.~\ref{fig:Pxt}(b), yield components
of the drift velocity, $v_i(z)$, which appear in Fig.~\ref{fig:vz}.
Fitting $v_i(z)$ to Eq.~(\ref{eq:vz})
for $v_{0,i}$, $H$, and $z_0$,
yields $H = 8.16 \pm 0.22$~\micron and fixes the 
lower wall's absolute position to within $\pm 0.13$~\micron.
Errors primarily reflect the experimental uncertainty
of $\pm 0.25$~\micron in the increment between tweezer locations.
Without loss of generality, we set $z_0 = 0$ and
measure starting heights $h = z - z_0$ from the lower glass surface.

\begin{figure}[t!]
  \begin{center}
    \includegraphics[width=3in]{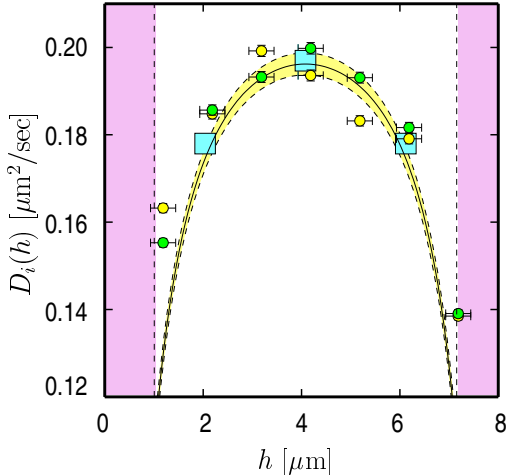}
    \caption{Height dependence of the in-plane diffusion coefficient
      for a Brownian sphere in a slit pore.  The solid curve
      results from the stokeslet approximation
      and is indistinguishable from
      the linear superposition approximation's prediction.
      Dashed curves show the range of predicted
      values due to the uncertainty in the sphere's radius.
      Squares show Fax\'en's predictions from Eq.~(\protect\ref{eq:faxen2}).
      }
    \label{fig:Dz}
  \end{center}
\end{figure}

The evolution of the distributions' widths over time,
shown in Fig.~\ref{fig:Pxt}(c),
can be interpreted with Eq.~(\ref{eq:stokeseinstein})
to obtain the drag-corrected self-diffusion coefficients $D_i(h)$
for each height.
Results are summarized in Fig.~\ref{fig:Dz}.
As expected, the in-plane diffusion coefficients measured for
motions along and transverse to the imposed flow agree
to within 5\% at all heights.
The sphere's diffusivity is strongly suppressed near either wall, 
and falls well below $D_0$ even along the midplane.

Fax\'en's result for the drag on a sphere near a single wall
provides a good starting point for interpreting these observations.
His result for the in-plane drag, $\gamma_1(h)$,
\begin{equation}
  \label{eq:onewall}
  \frac{\gamma_0}{\gamma_1(h)} = 
  1 - \frac{9}{16} \, \frac{a}{h} + 
    {\cal O} \left(\frac{a^3}{h^3}\right)
\end{equation}
has been
verified to great accuracy using photonic force microscopy
in a thick sample cell \cite{pralle98}.
Oseen suggested in 1927 that 
the drag due to two parallel walls could be approximated
by a sum of two such single-wall contributions \cite{oseen27}:
\begin{equation}
  \label{eq:oseen}
  \gamma_2(h) \approx \gamma_0 + [\gamma_1(h) - \gamma_0] + 
  [\gamma_1(H-h) - \gamma_0].
\end{equation}
Even though this linear superposition approximation violates 
boundary conditions at both bounding surfaces,
the corresponding prediction for the in-plane diffusivity,
$D_2(h) = k_B T / \gamma_2(h)$, agrees quite well
with our measurements, with no adjustable parameters, 
as can be seen in Fig.~\ref{fig:Dz}.
Higher-order corrections \cite{happel91} to $\gamma_1(h)$
yield negligibly small corrections to $D_2(h)$ and
so are not presented.

Faucheux and Libchaber adopted the linear superposition
approximation, averaged
the resulting in-plane diffusivity over $h$,
and obtained reasonably good agreement
with $h$-averaged measurements for various sphere radii
\cite{faucheux94}.
Lin, Yu and Rice further
demonstrated linear superposition's accuracy
at $h = H/2$ by measuring 
a sphere's in-plane and out-of-plane
diffusivity near the midplanes of slit pores of various widths 
\cite{lin00}.

Regardless of its empirical success, the linear superposition
approximation fails to satisfy boundary conditions and so
cannot yield accurate predictions under all
circumstances.
Lobry and Ostrowsky attempted to remedy its
shortcomings by accounting
for flows' reflections off bounding surfaces \cite{lobry96}.
Considering such corrections up to third order, they
obtained results which compare rather poorly
with measurements
by Lin, Yu and Rice \cite{lin00}.

Fax\'en managed to
calculate two-wall drag coefficients for the particularly
symmetric arrangements $h = H/4$ and $h = H/2$
\cite{happel91}.
The in-plane results,
\begin{equation}
  \label{eq:faxen2}
  \frac{\gamma_0}{\gamma_F(h)} = 
  \begin{cases}
    1 - 0.6526 \, \frac{a}{h} + 0.1475 \, \frac{a^3}{h^3} \\
   \qquad - 0.131 \, \frac{a^4}{h^4}
   - 0.0644 \, \frac{a^5}{h^5},
   & h = \frac{H}{4} \\
   1 - 1.004 \, \frac{a}{h} + 0.418 \, \frac{a^3}{h^3} \\
   \qquad + 0.21 \, \frac{a^4}{h^4}
   - 0.169 \, \frac{a^5}{h^5}, 
   & h = \frac{H}{2}
  \end{cases}
\end{equation}
appear as squares in Fig.~\ref{fig:Dz}
and agree well with both Oseen's approximation and also with our
measurements.
Unfortunately, Fax\'en's method does not work for other configurations.

Another approach, due to Blake \cite{blake71}, takes advantage
of an analogy between hydrodynamics and electrostatics.
Rather than attempting to satisfy no-slip boundary conditions at a bounding
wall directly, Blake introduced the notion of a hydrodynamic
image whose flow field exactly cancels the particle's
on the boundary.
Uniqueness of the Stokes equation's solutions guarantees that the image
solution also solves the original problem.

Blake's treatment yields the Green's function for a particle's
flow field bounded by a single surface.
Accounting for a second parallel surface involves 
adding not only the particle's 
hydrodynamic image in the new surface, 
but also an infinite series of images of the images.
Liron and Mochon \cite{liron76} summed this series explicitly for a point
force acting on the confined fluid, derived the additional
terms needed to cancel residual flows on the surfaces,
and thereby obtained the Green's function, $G(h \hat z - \vec r,H)$,
for the flow at $\vec r$ due to a disturbance a height $h$ above the
lower wall in a slit pore of width $H$.
Integrating their result over the sphere's surface would yield 
an accurate though unwieldy expression for $\gamma(h)$.

A much simpler solution can be obtained when far field contributions
dominate the flows at the bounding surfaces.
This is the case when the sphere is much
smaller than the gap between the walls, $a \ll H$.
In this limit, the flow field due to the sphere's hydrodynamic
images is well approximated by the image flow for a point force.
Thus, the drag on a sphere centered
a height $h$ above the lower wall
may be approximated by
\begin{equation}
  \label{eq:stokeslet}
  \gamma^{-1}(h) \approx \gamma_0^{-1} + [G(0,H) - G_S(0)],
\end{equation}
where $G_S(0)$ is the singular contribution due to a unit point force
at the sphere's location, so that the term in square brackets is the
flow at the sphere's location due to the images alone.
More generally, $G_S(\vec r)$ is the Green's function in an unbounded fluid
and is known as a stokeslet.
Eq.~(\ref{eq:stokeslet}), then, is an example of a stokeslet approximation.
Evaluating Eq.~(\ref{eq:stokeslet})
using the expressions for $G(\vec r)$ and $G_S(\vec r)$
in Equations (5), (15), (32) and (33) of Ref.~\cite{liron76} 
yields the in-plane diffusion coefficient
\begin{multline}
  \frac{D(h)}{D_0} = 1 + \frac{3}{4} \frac{a}{H} \, \int_0^\infty
  \left[A_0 \left( \lambda, \frac{h}{H} \right) \right.\\\left.
    - \lambda^2 \, A_1  \left( \lambda, \frac{h}{H} \right)
    - 1 \right] \, d\lambda,
  \label{eq:liron}
\end{multline}
where
\begin{equation}
  \label{eq:liron2}
  A_0(\lambda,\eta) = 2 \, \frac{
    \sinh(\lambda \eta) \, \sinh[ \lambda(1-\eta)]}{\sinh(\lambda)}
\end{equation}
is the contribution due to a stokeslet at the sphere's position 
together with
an infinite series of image stokeslets, and
\begin{multline}
  \label{eq:liron3}
  A_1(\lambda,\eta) = \frac{1}{\sinh^2(\lambda) - \lambda^2} \times \\
  \Biggl\{   
  \eta^2 \sinh(\lambda) \cosh[\lambda(1-2\eta)] - \sinh(2 \lambda \eta) + \\
  \lambda \eta^2 \sinh^2[\lambda(1-\eta)] +
  \coth(\lambda) \sinh^2(\lambda \eta) -  \\
  \frac{\lambda}{4} \left[ \frac{(2-\eta) \sinh(\lambda \eta) -\eta \sinh[\lambda(2-\eta)]}{\sinh(\lambda)} \right]^2
   \Biggr\}
\end{multline}
enforces the no-flow boundary conditions at the walls.
The numerically evaluated result appears as the 
solid curve in Fig.~\ref{fig:Dz}.

Remarkably, the predictions of Eqs.~(\ref{eq:liron}), (\ref{eq:liron2})
and (\ref{eq:liron3})
are indistinguishable from
Oseen's superposition approximation for our experimental conditions,
and both agree quantitatively with Fax\'en's fifth-order
results, Eq.~(\ref{eq:faxen2}).
Further comparison with Eq.~(\ref{eq:faxen2})
reveals that the stokeslet
approximation becomes increasingly accurate for larger wall separations,
while the linear superposition approximation fares less well.
For smaller separations, on the other hand, 
the stokeslet approximation becomes less accurate.
Linear superposition performs surprisingly well at small separations, 
by contrast, particularly
if higher order corrections to Eq.~(\ref{eq:onewall}) are included.

Despite the apparent complexity of Eqs.~(\ref{eq:liron}),
(\ref{eq:liron2}) and (\ref{eq:liron3})
when compared with the linear superposition approximation, 
stokeslet analysis has the appeal of scalability.  
Under conditions for which the stokeslet approximation is valid,
contributions to the diffusivity tensor for a system of spheres can
be combined in a straightforward manner \cite{brenner99},
as in Eq.~(\ref{eq:stokeslet}).
We recently demonstrated this approach's utility in
the comparatively simple case of two spheres near one wall 
for which linear superposition yields inaccurate predictions
\cite{dufresne00}.
Further demonstrating its accuracy for
the much more challenging case of two-wall confinement lends confidence
in its potential for systematically treating
many-body colloidal hydrodynamics.

We are grateful to Todd Squires and Michael Brenner for introducing us
to stokeslet analysis in general and to Ref.~\cite{liron76} in particular.
This research was supported primarily by the National Science Foundation
through Award Number DMR-978031.  DA was supported in part by The
University of Chicago MRSEC REU Program through Award Number DMR-980595. 
Additional support was
provided by the David and Lucile Packard Foundation.

\vspace{-2mm}

\end{document}